# 3D Lymphoma Segmentation on PET/CT Images via Multi-Scale Information Fusion with Cross-Attention

Huan Huang[1]·Liheng Qiu[2]·Shenmiao Yang[3]·Longxi Li[1]·Jiaofen Nan[1]·Yanting Li[1]·Chuang Han[1]·Fubao Zhu[1*]·Chen Zhao[4*]·Weihua Zhou[5,6]

[1] School of Computer Science and Technology, Zhengzhou University of Light Industry, Zhengzhou 450002, Henan, China
[2] Peking University Peoples' Hospital, Peking University Institute of Hematology, Beijing, China
[3] Department of Nuclear Medicine, Peking University Peoples' Hospital, Beijing, China
[4] Department of Computer Science, Kennesaw State University, Marietta, GA, USA
[5] Department of Applied Computing, Michigan Technological University, Houghton, MI, USA
[6] Center for Biocomputing and Digital Health, Institute of Computing and Cybersystems, and Health Research Institute, Michigan Technological University, Houghton, MI, USA

Huan Huang and Liheng Qiu contributed equally.

[*] Correspondence:
Fubao Zhu
Email address: fbzhu@zzuli.edu.cn
Mailing address: School of Computer Science and Technology, Zhengzhou University of Light Industry, Zhengzhou 450002, Henan, China
Chen Zhao
Email address: czhao4@kennesaw.edu
Mailing address: 680 Arntson Dr, Atrium BLDG, Marietta, GA 30060, USA

**Abstract**

**Background:** Accurate segmentation of diffuse large B-cell lymphoma (DLBCL) lesions is challenging due to their complex patterns in medical imaging.

**Objective**: This study aims to develop a precise segmentation method for DLBCL using 18F-Fluorodeoxyglucose (FDG) positron emission tomography (PET) and computed tomography (CT) images.

**Methods:** We propose a 3D dual-branch encoder segmentation method using shifted window transformers and a Multi-Scale Information Fusion (MSIF) module. To enhance feature integration, the MSIF module performs multi-scale feature fusion using cross-attention mechanisms with a shifted window framework. A gated neural network within the MSIF module dynamically balances the contributions from each modality. The model was optimized using the Dice Similarity Coefficient (DSC) loss function. Additionally, we computed the total metabolic tumor volume (TMTV) and performed statistical analyses.

**Results**: The model was trained and validated on a dataset of 165 DLBCL patients using 5-fold cross-validation, achieving a DSC of 0.7512. Statistical analysis showed a significant improvement over comparative methods ($p < 0.05$). Additionally, a Pearson correlation coefficient of 0.91 and an $R^2$ of 0.89 were observed when comparing manual annotations to segmentation results for TMTV measurement.

**Conclusion**: This study presents an effective automatic segmentation method for DLBCL that leverages the complementary strengths of PET and CT imaging. Our method has the potential to improve diagnostic interpretations and assist in treatment planning for DLBCL patients.

# 1. Introduction

Diffuse large B-cell lymphoma (DLBCL) is a common subtype of non-Hodgkin lymphoma[1, 2]. Approximately two-thirds of DLBCL patients can be cured with Rituximab, Cyclophosphamide, Doxorubicin, Vincristine, and Prednisone (R-CHOP) like chemoimmunotherapy[3]. In clinical practice, 18F-Fluorodeoxyglucose (FDG) positron emission tomography (PET) and computed tomography (CT) are commonly used for DLBCL staging and treatment response assessment[4-6]. Integrating PET and CT imaging is crucial for lymphoma segmentation, as it combines metabolic activity from PET with anatomical details from CT[7].

Moreover, total metabolic tumor volume (TMTV), which quantifies the metabolic activity of tumors, is a key prognostic biomarker for DLBCL[8]. Accurate lymphoma segmentation is essential for determining TMTV, but manual delineation is both time-consuming and subjective. Recent advances in deep learning have led to the development of automated segmentation methods, providing greater consistency and accuracy[9].

Traditional lymphoma segmentation methods, such as thresholding and region growing, have inherent limitations. Thresholding, while straightforward, lacks adaptability[10], especially when image conditions cause lymphoma and normal tissue to appear with similar gray values. Region growing is highly dependent on initial seed points[11], which require careful selection to handle the diverse shapes and sizes of lymphoma. However, recent progress in deep learning has demonstrated its potential to address these challenges, offering automated solutions that significantly improve segmentation accuracy and consistency.

Li et al.[12] proposed an end-to-end network for semi-supervised lymphoma segmentation, achieving a Dice similarity coefficient (DSC) of 0.72 using PET/CT data from 80 lymphoma cases. Yuan et al.[13] introduced a dual-branch encoder network for lymphoma segmentation, achieving a DSC of 0.73 on 45 DLBCL patients. Blanc-Durand et al.[14] achieved a DSC of 0.73 using their 3D U-Net, trained and validated on PET/CT data from 639 DLBCL patients, with 94 cases reserved for testing. Yousefirizi et al.[15] proposed a cascaded approach for automated tumor delineation in lymphoma involving 1418 PET/CT scans from multiple centers. This approach combined multi-resolution 3D U-Nets and model ensembling, achieving an average DSC of 0.68 on internal test data and 0.66 on external multi-site data. While these methods demonstrated notable results, they did not fully leverage the complementary information from PET and CT multimodal data. Furthermore, the limited receptive field of convolutional neural networks (CNNs) poses challenges in capturing both global and local information, particularly in small lesion areas[16-18].

This study addresses these gaps and makes several key contributions to the field of lymphoma diagnosis:

1) **Multimodal feature extraction:** We introduce a novel dual-branch encoder based on the Swin Transformer[19, 20], which effectively captures global features from both PET and CT modalities, overcoming the limitations of traditional CNNs in handling receptive fields.

2) **Multimodal feature fusion:** A Multi-Scale Information Fusion (MSIF) module is developed, integrating multi-scale feature fusion with a cross-attention mechanism using shifted windows. This module enhances the interaction between fine-grained features from PET and CT, facilitating efficient information exchange across modalities.
3) **Applying lymphoma segmentation to TMTV calculation:** In addition to segmentation, this study focuses on the evaluation of TMTV, a key metric for evaluating treatment effectiveness and prognosis.

## 2. Method and Material
### 2.1 Dataset

This study utilized 165 PET/CT scan datasets from patients clinically diagnosed with DLBCL, provided by Peking University People's Hospital. The PET/CT scans were acquired using a Discovery VCT PET/CT scanner (GE Healthcare, Milwaukee, Wisconsin, USA), and image reconstruction was performed using the Ordered Subset Expectation Maximization method. Ground truth volumes of interest (VOI) were manually segmented on PET images by experienced nuclear medicine experts to ensure accuracy. PET images were acquired one hour after the intravenous injection of 18F-FDG to capture optimal metabolic activity

The size of CT images is 512×512, with a pixel spacing of 0.98 mm×0.98 mm per pixel, while the size of PET images is 128×128, with a pixel size of 5.47 mm×5.47 mm per pixel. Both PET and CT images were reconstructed with an identical number of slices and a slice thickness of 3.27 mm.

### 2.2 Data preprocessing

Rigid-body registration was used to align the PET and CT volumes into the same coordinate space, a standard practice in PET/CT segmentation[21, 22]. The PET images were upsampled to a target size of 256×256 using bicubic interpolation, while the CT images were downsampled to the same size. Subsequently, all slices were cropped to 224×224 pixels, removing peripheral regions irrelevant to segmentation.

Based on clinical recommendations, the CT image window width was adjusted to 400 Hounsfield units (HU), and the window level to 40 HU. For the PET images, we normalized the pixel values using Body Weight-Corrected Standard Uptake Value ($SUV_{BW}$), a semi-quantitative measure commonly used to evaluate FDG uptake[6, 23], as defined in Eq. (1).

$$SUV_{BW} = \frac{RS \times PV + RI}{A \times e^{\frac{-0.693}{T_{\frac{1}{2}}} \times (t_1 - t_0)} / (W \times 1000)} \qquad (1)$$

where $PV$ represents the pixel value of the PET slices, $RS$ and $RI$ denote the rescale slope and intercept of PET imaging, respectively, $A$ is the total dose of the radioactive isotope, $T_{\frac{1}{2}}$ is the radioactive isotope half-life of 18F-FDG, $t_0$ is the start time of the radiopharmaceutical injection, $t_1$ is the image acquisition time, and $W$ is the weight of the patient.

### 2.3 Network architecture

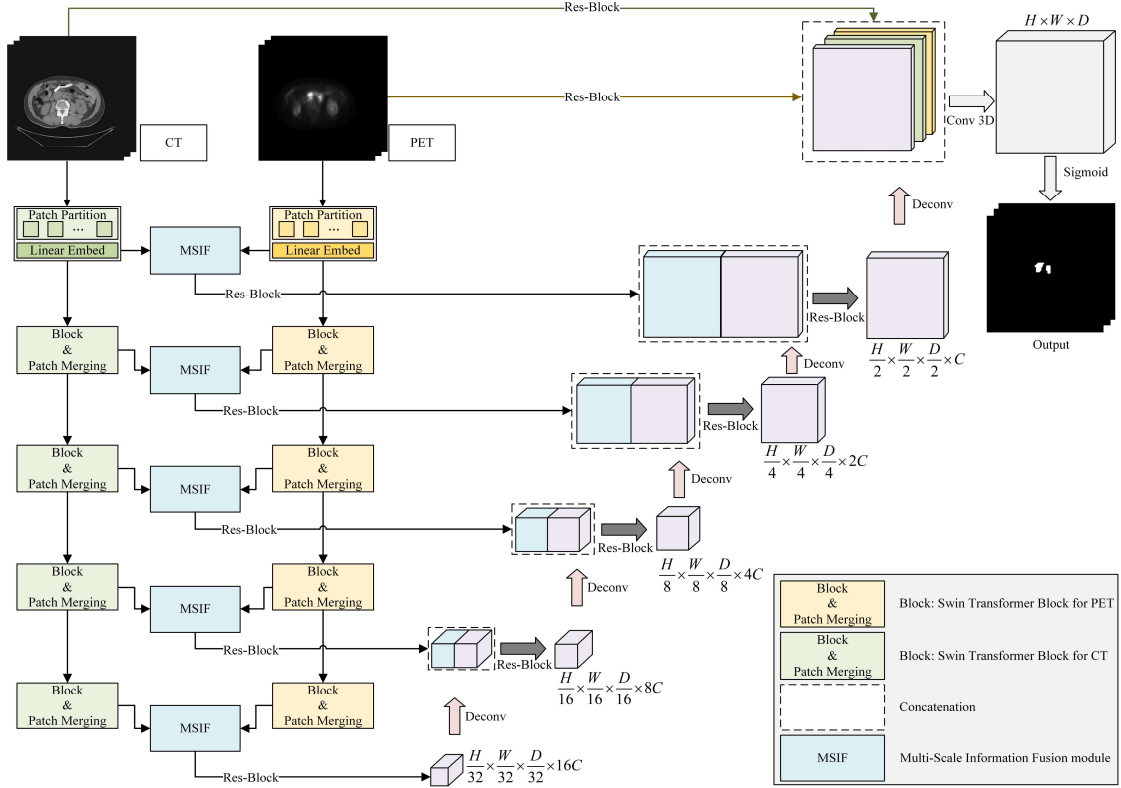

Fig. 1: The network consists of two independent encoder branches, which process PET and CT modality images separately. Each modality's image is first passed through a linear embedding layer to generate intermediate feature maps. These feature maps are then processed through four stages of Swin Transformer Blocks and Patch Merging, resulting in multi-scale feature maps with progressively reduced resolutions. At each scale, the feature maps from both modalities are fused using the MSIF module for multi-scale information integration and then fed into the decoder. The decoder upscales the feature maps using skip connections and generates the final segmentation results after applying a Sigmoid activation function.

### 2.3.1 Encoders

Figure 1 illustrates the proposed network architecture, which consists of two independent encoders designed to fully extract raw features from each modality. Before the first stage, a 3D convolutional layer with a stride equal to the patch size is applied to each modality's 3D volume $X_{modal\_n} \in R^{H \times W \times D \times C}$, where $H$, $W$, and $D$ represent the dimensions of the feature maps, and $C$ denotes the number of channels. By using a stride equal to the patch size, the original voxel images are divided into non-overlapping patches of size $H' \times W' \times D'$ ($2 \times 2 \times 2$) and intermediate feature maps with reduced dimensions $\frac{H}{2} \times \frac{W}{2} \times \frac{D}{2} \times C$ are generated, where the number of output channels is set to $C = 16$.

These patches are projected into an embedding space, forming tokens with a dimension of $D_{embed} = 24$. These tokens are organized into a sequence $S_{modal\_n} \in R^{N \times D_{embed}}$, where $N = \frac{H}{H'} \times \frac{W}{W'} \times \frac{D}{D'}$ representing the number of tokens.

Each encoding branch follows the structure of the Swin-UnetR[20] encoder, which consists of 4 stages. Each stage contains 2 transformer blocks, resulting in a total of 8 layers ($L = 8$). At the end of each stage, a patch merging layer is applied to halve the resolution of the feature maps, resulting in feature maps with resolutions: $\frac{H}{4} \times \frac{W}{4} \times \frac{D}{4} \times 2C$, $\frac{H}{8} \times \frac{W}{8} \times \frac{D}{8} \times 4C$, $\frac{H}{16} \times \frac{W}{16} \times \frac{D}{16} \times 8C$, and $\frac{H}{32} \times \frac{W}{32} \times \frac{D}{32} \times 16C$, respectively.

These multi-scale feature maps are then utilized for cross-modal feature fusion through the MSIF module, enabling the model to effectively capture multi-scale cross-modal features. To minimize information loss, both the original feature map embeddings and the fused feature maps generated at each stage are fed into the decoder via skip connections, ensuring that critical information is preserved throughout the encoding and decoding processes.

### 2.3.2 Multi-Scale Information Fusion

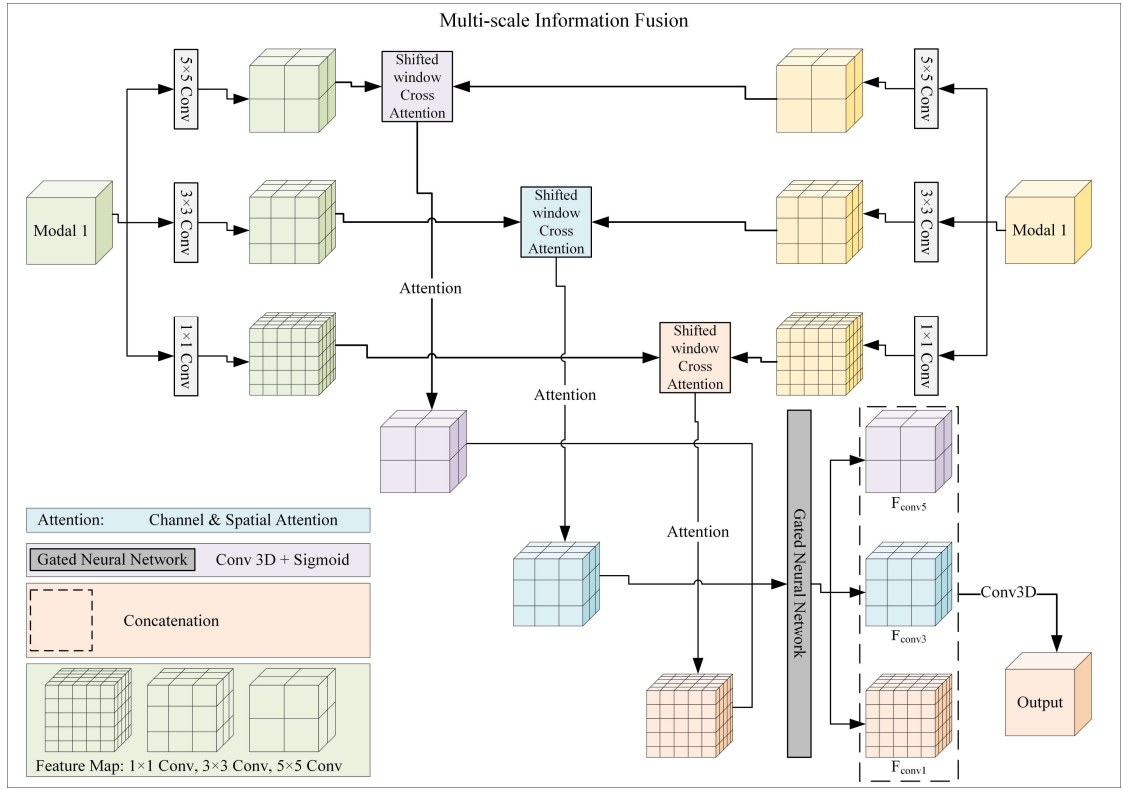

Fig. 2: The structure of MSIF. The MSIF module fuses multi-scale feature maps from both modalities using convolutional layers with different kernel sizes, followed by shifted window cross-attention and channel-spatial attention mechanisms. The final output feature maps are processed through a gating network and concatenated before being fused in a 3D convolutional layer.

To achieve finer-grained fusion of features from the two modalities, we developed a multi-scale cross-modal feature fusion method, called the MSIF module. As shown in Figure 2, this module receives intermediate feature maps generated by both modalities and then applies convolutional layers with different kernel sizes (set to 1, 3, and 5 in this study) to produce the multi-scale features. Once the feature maps of different scales are calculated, the corresponding scaled feature maps from both modalities are fused using the shifted window cross-attention module.

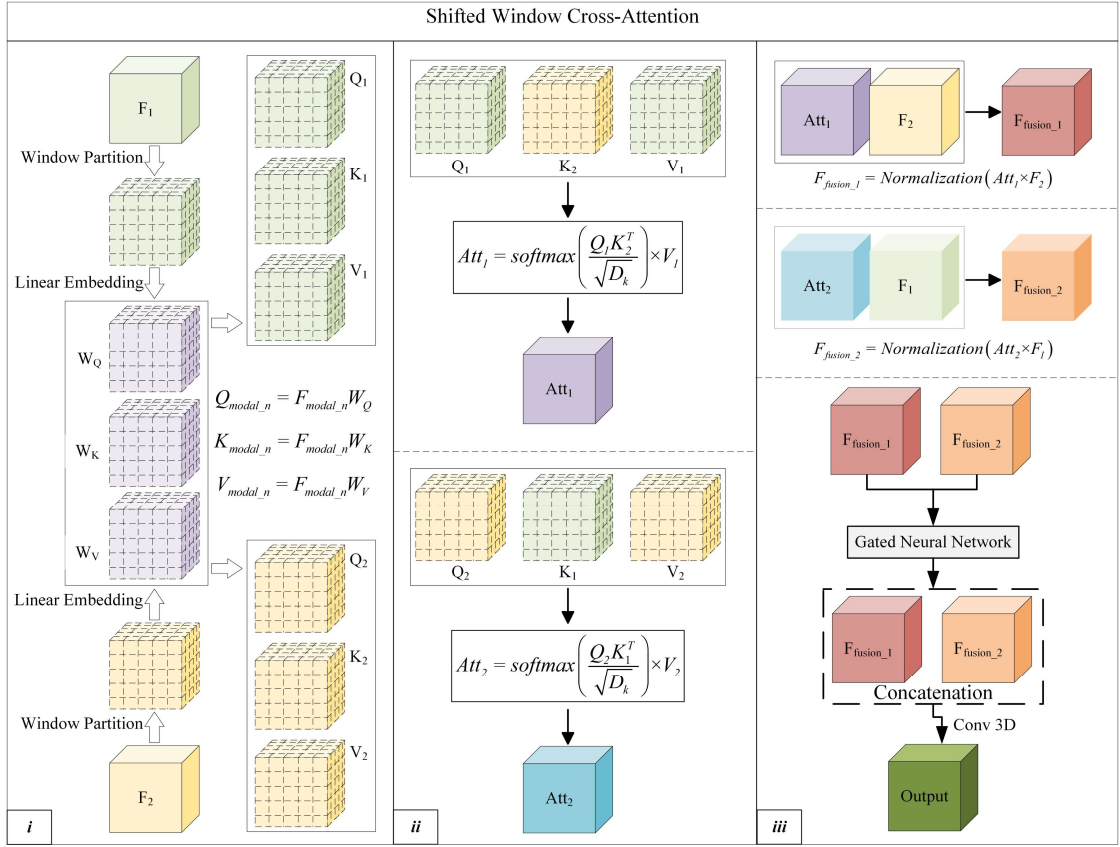

Fig. 3 The architecture of the shifted window cross-attention. (***i***) $Q_1, K_1$, and $V_1$ are obtained by multiplying the intermediate feature map from the first modality with the learnable weight matrices $W_Q$, $W_K$, and $W_V$. Similarly, $Q_2$, $K_2$, and $V_2$ are derived from the intermediate feature map of the second modality using the same weight matrices. (***ii***) The cross-attention mechanism. (***iii***) Cross modal feature fusion.

To minimize computational costs of cross-attention, we implement a 3D shifted window mechanism. Initially, each modality's 3D input image $F_{modal\_n}$ is split into non-overlapping patches of size $M \times M \times M (7 \times 7 \times 7)$. As shown in Figure 3 (***i***), we compute queries, keys, and values for each patch as follows:

$$Q_{modal\_n}^l = F_{modal\_n}^l W_Q \tag{2}$$

$$K_{modal\_n}^l = F_{modal\_n}^l W_K \tag{3}$$

$$V_{modal\_n}^l = F_{modal\_n}^l W_V \tag{4}$$

where $l$ is the layer of Swin Transformer. $W_Q, W_K, W_V \in R^{D_f \times D_q}$ are the weight matrices, where $D_f$ is the feature dimension, and $D_q$ is the dimension of the queries and keys.

As shown in Figure 3 (***ii***), cross-modal attention is then computed as shown in Eq. 5:

$$Att_1^l = softmax\left(\frac{Q_1^l \left(K_2^l\right)^T}{\sqrt{D_k}}\right)V_1^l, \quad Att_2^l = softmax\left(\frac{Q_2^l \left(K_1^l\right)^T}{\sqrt{D_k}}\right)V_2^l \qquad (5)$$

where $D_k$ is the dimension of the keys and queries. Empirically, we set $D_q = D_k$. $T$ denotes the matrix transpose, necessary for matching the dimensions of queries and keys during the dot-product operation.

At each layer $l$ the image is partitioned into windows of size $M \times M \times M$, In the subsequent layer $l+1$, the windows are shifted by $\left[\frac{M}{2}, \frac{M}{2}, \frac{M}{2}\right]$ voxels, enabling interaction between adjacent windows and reducing redundant calculations. The outputs for layers $l$ and $l+1$ are computed using Eqs. 6 to 9:

$$\hat{A}_{modal\_n}^l = W\text{-}MSA\left(LN\left(A_{modal\_n}^{l-1}\right)\right) + \hat{A}_{modal\_n}^{l-1} \qquad (6)$$

$$A_{modal\_n}^l = MLP\left(LN\left(\hat{A}_{modal\_n}^l\right)\right) + \hat{A}_{modal\_n}^l \qquad (7)$$

$$\hat{A}_{modal\_n}^{l+1} = SW\text{-}MSA\left(LN\left(A_{modal\_n}^l\right)\right) + A_{modal\_n}^l \qquad (8)$$

$$A_{modal\_n}^{l+1} = MLP\left(LN\left(\hat{A}_{modal\_n}^{l+1}\right)\right) + \hat{A}_{modal\_n}^{l+1} \qquad (9)$$

In these equations, $W-MSA$ and $SW-MSA$ stand for regular and shifted window multi-head self-attention modules, respectively. The regular multi-head self-attention $(W-MSA)$ operates within fixed windows, while the shifted window version $(SW-MSA)$ shifts the windows by half their size in each dimension to enable interaction between adjacent windows. This mechanism reduces redundant calculations and enhances feature extraction across the entire input[19]. $LN$ denotes layer normalization, and $MLP$ refers to the multi-layer perceptron.

Figure 3 (***iii***) illustrates the fusion step, described as in Eqs. 10 to 12:

$$F_{fusion\_1} = Normalization\left(Att_1 \times F_2\right) \qquad (10)$$

$$F_{fusion\_2} = Normalization\left(Att_2 \times F_1\right) \qquad (11)$$

$$F_{out} = Conv(gate(F_{fusion\_1}) + gate(F_{fusion\_2})) \qquad (12)$$

where $Att_{modal\_n}$ represents each modality's output from the shifted window cross-attention, and *gate* refers to a gating mechanism comprising a convolutional layer followed by a Sigmoid activation function. The Sigmoid function is used to scale the fusion weights between $[0, 1]$, providing a smooth transition and allowing for effective control over the influence of each modality in the final feature fusion process.

After the cross-modal feature fusion, the feature maps at each scale are sequentially processed by a channel attention module and a spatial attention module for feature enhancement. The enhanced feature maps are then fed into a lightweight gated neural network, which assigns weights to the features based on their contribution to segmentation performance. Finally, these weighted multi-scale features are further fused to achieve comprehensive multi-scale and multimodal feature integration. This is described in Eq. 13:

$$F_{final} = Conv(gate(F_{conv1}) + gate(F_{conv3}) + gate(F_{conv5})) \tag{13}$$

where $F_{conv1}, F_{conv3}$, and $F_{conv5}$ are the outputs at different scales from the shifted window cross-attention module.

### 2.3.3 Decoder

The decoder follows a U-shaped architecture. In the deepest layer of the encoder, the fused feature maps are resized to $\frac{H}{32} \times \frac{W}{32} \times \frac{D}{32}$ and then passed through a residual block, which includes 3D convolutions and instance normalization layers. Next, these feature maps are processed through a series of upsampling blocks, each containing a residual block. These are combined with the corresponding feature maps from the encoder via skip connections. After combining, the feature maps are passed through another residual block to further refine the segmentation. Each upsampling block consists of a deconvolutional layer, a normalization layer, and a ReLU activation function, with an upsampling factor of 2. Finally, the segmentation probability maps are calculated through a $1 \times 1 \times 1$ 3D convolution layer followed by a Sigmoid activation function.

### 2.4 Segmentation evaluation criteria

We employ the following evaluation metrics to assess the segmentation model's performance:

**Dice Similarity Coefficient (DSC)**: DSC is commonly used to measure segmentation accuracy, as shown in Eq. 14:

$$DSC = \frac{2 \times TP}{2 \times TP + FP + FN} \tag{14}$$

where $TP$ (True Positives) represents the number of correctly classified positive samples, $FP$ (False Positives) represents the number of misclassified positive samples, $FN$ (False Negatives) represents the number of misclassified negative samples, and $TN$ (True Negatives) represents the number of correctly classified negative samples.

**Sensitivity** is calculated as follows:

$$Sensitivity = \frac{TP}{TP + FN} \tag{15}$$

**Precision** is calculated as follows:

$$Precision = \frac{TP}{TP + FP} \tag{16}$$

### 2.5 Implementation and experiments

We implemented the network using PyTorch 1.10.0[24]. Training was conducted on an Ubuntu 16.04 server equipped with a Tesla V100 GPU. For optimization, we employed the Adam optimizer[25] with a dynamic learning rate. Initially, the learning rate was set to 0.001, with $\beta 1$ and $\beta 2$ empirically set to 0.9 and 0.999, respectively. We employed DSC loss to penalize the difference between the predicted segmentation maps and the ground truth of lymphoma, as defined in Eq. 17:

$$L = 1 - \frac{2\sum_{i=1}^{N}(p_i \times g_i) + \varepsilon}{\sum_{i=1}^{N}p_i + \sum_{i=1}^{N}g_i + \varepsilon} \quad (17)$$

Where $p_i$ represents the predicted value for the $i$-th pixel, $g_i$ represents the ground truth value for the $i$-th pixel, $N$ is the total number of pixels, and $\varepsilon$ is a small scalar to avoid division by zero.

To demonstrate the effectiveness of our method, we compared it with various state-of-the-art (SOTA) methods on the lymphoma segmentation dataset, including UnetR[26], Swin-UnetR[20], Att-Unet[27], Unet++[28], SegResNet[29], and SwinCross[30]. We applied 5-fold cross-validation, with each training set consisting of 132 subjects and each test set consisting of 33 subjects. To meet the input requirements of the 3D network and avoid GPU memory limitations, we employed a sliding window technique. Specifically, 32 consecutive slices were extracted per batch to form a 3D volume. The training and validation sets were divided based on individual patient data to ensure the model's generalization ability.

**2.6 TMTV calculation**

The TMTV is calculated by summing the volume of voxels in both the binarized ground truth and the predicted masks (where the prediction probability exceeds 0.5). This sum is then multiplied by the voxel volume to obtain the TMTV[31]. The formula is defined in Eq. 18.

$$TMTV = \sum_{i=1}^{S} V_i \times V_{voxel} \quad (18)$$

where $V_i$ represents the $i$-th voxel in the binarized mask, $V_{voxel}$ is the volume of a single voxel, and $S$ is the total number of voxels.

To evaluate the relationship between the calculated TMTV (cTMTV) and the ground truth TMTV (gTMTV), we conduct linear regression and calculate the coefficient of determination $R^2$. We also assess the correlation between the two TMTV measures using Pearson's correlation coefficient. Additionally, Bland-Altman analysis is used to evaluate the agreement between the cTMTV and gTMTV. Consistent with the segmentation task, we analyze the results from each fold of the five-fold cross-validation to ensure the reliability and stability of our findings.

# 3 Results
## 3.1 Segmentation results

**Table 1:** Results of different methods in lymphoma segmentation.

| Method | DSC | Sensitivity | Precision |
| --- | --- | --- | --- |
| UnetR | 0.7107±0.0178 ** | 0.7608±0.0128 ** | 0.6686±0.0298 ** |
| SegResNet | 0.7223±0.0146 ** | 0.7175±0.0466 | 0.7289±0.0125 ** |
| Swin-UnetR | 0.7271±0.0163 ** | **0.7659±0.0123 **** | 0.7041±0.0246 ** |
| SwinCross | 0.7414±0.0209 ** | 0.7405±0.0213 ** | 0.7432±0.0176 ** |
| Unet++ | 0.7446±0.0129 * | 0.7322±0.0072 ** | 0.7577±0.0137 ** |
| Att-Unet | 0.7463±0.0113 ** | 0.7622±0.0075 ** | 0.7314±0.0179 ** |
| Ours | **0.7512±0.0078** | 0.7548±0.0063 | **0.7611±0.0078** |

The best metric is shown in bold. All p-values are below 0.1 compared to our method, where * indicates p < 0.05, and ** indicates p < 0.01.

Table 1 shows the performance of various methods in lymphoma segmentation. Our model achieves the highest DSC (0.7512) and precision (0.7611), demonstrating its significant advantages in overall performance and false positive reduction. While our sensitivity score (0.7548) is slightly lower than Swin-UnetR's (0.7659), it remains competitive, indicating that our approach effectively balances different metrics for accurate and reliable tumor segmentation.

UnetR performs better in sensitivity but has a notably lower DSC due to its simpler architecture, which may not capture complex feature interactions effectively. Swin-UnetR, combining Swin Transformer with a U-shaped structure, achieves the highest sensitivity but falls short in segmentation and precision compared to our model, suggesting limitations in precise segmentation. SwinCross, an enhancement of Swin-UnetR with a dual-branch encoder and cross-attention mechanism, performs well but does not leverage multi-scale information as effectively as our model. Unet++ and SegResNet show strong DSC and precision but have lower sensitivity, indicating potential difficulties in identifying all tumor regions. Despite higher scores in some metrics, Att-Unet is outperformed by our model across all measures, suggesting its attention mechanisms are less effective at capturing multi-scale features.

To evaluate our method's stability under different conditions, we used box plots to display the distribution of DSC, sensitivity, and precision across five-fold cross-validation. Figure 4 shows that our method has a more concentrated distribution with less variability, indicating higher stability across experiments.

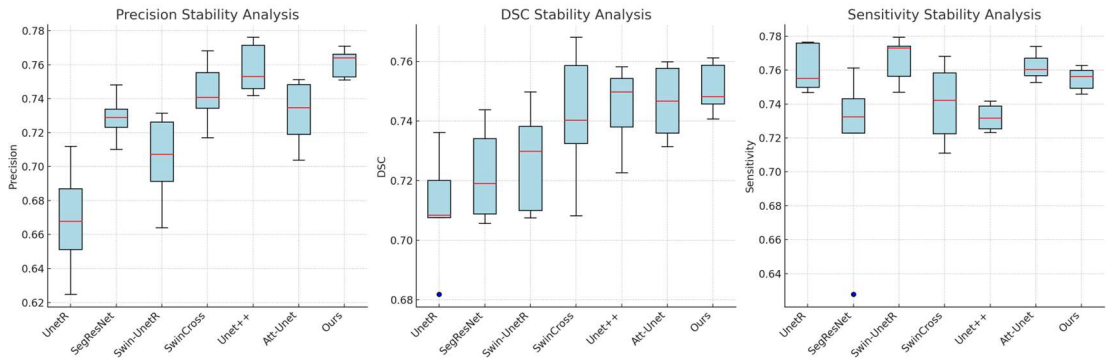

Figure 4. Stability analysis using box plots: This figure presents the ranges of precision, DSC, and sensitivity across five cross-validation folds for different models. Each box plot represents the performance of a model, with the following components: The box shows the 1st quartile (lower boundary), median (red line), and 3rd quartile (upper boundary), indicating the middle distribution of the data. The whiskers represent the range of the data, excluding outliers. Blue dots indicate outliers, which are observations that significantly deviate from the other data points.

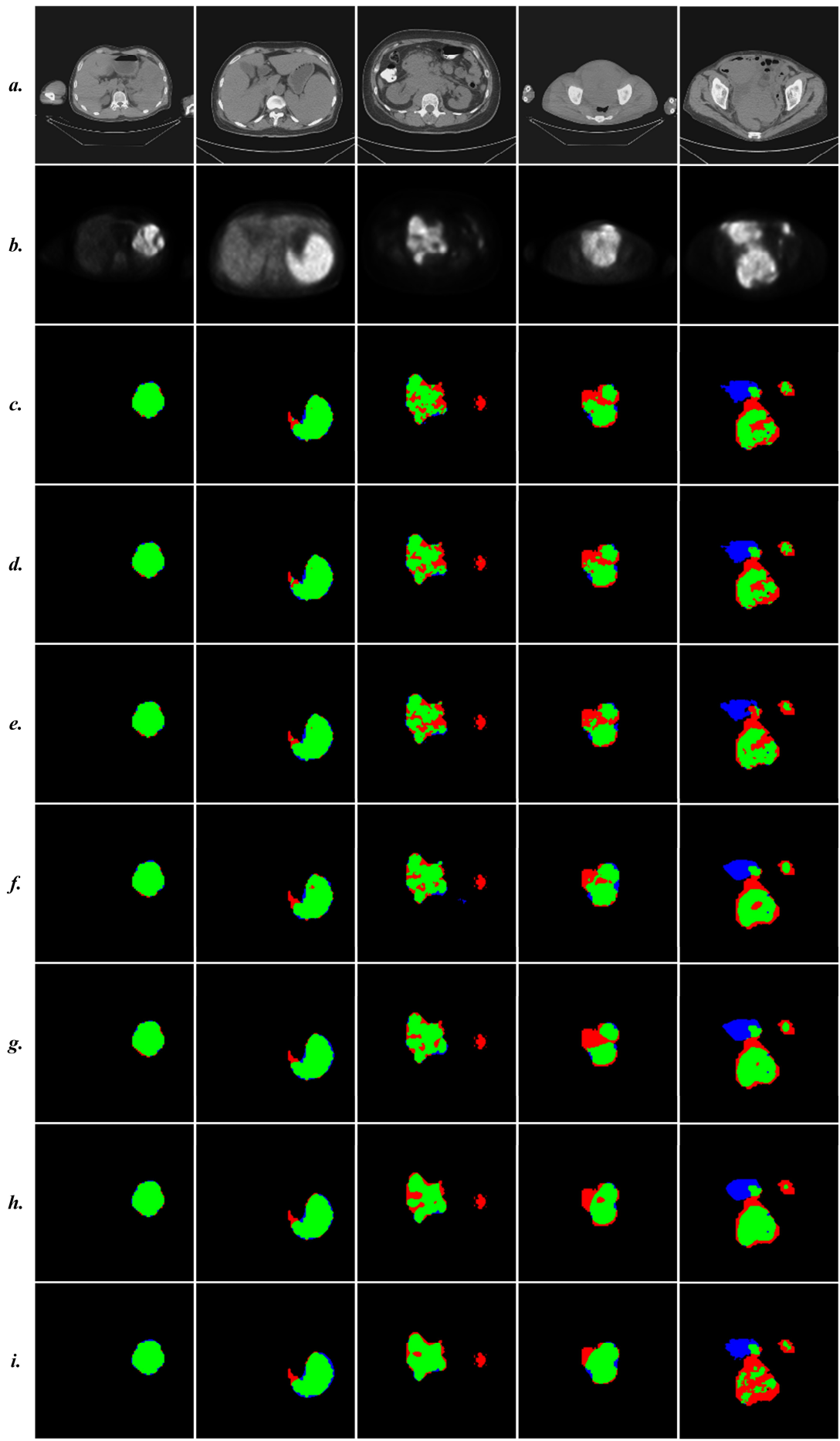

Fig. 5. Difference maps of segmentation results compared with ground truth. The green, red, and blue regions represent true positive, false negative, and false positive pixels, respectively. (***a***). CT images, (***b***). PET images, Difference maps generated by (***c***). our method. (***d***). Att-Unet. (***e***). Unet++. (***f***). SwinCross. (***g***). Swin-UnetR. (***h***). SegResNet. (***i***). UnetR.

To provide a more intuitive comparison of different methods in the segmentation task, we visualized the segmentation results using difference maps. Figure 5 highlights the effectiveness of different methods, particularly in smaller lesion regions. In areas with complex shapes or blurred edges, our model more accurately reproduces the ground truth.

## 3.2 Results of TMTV

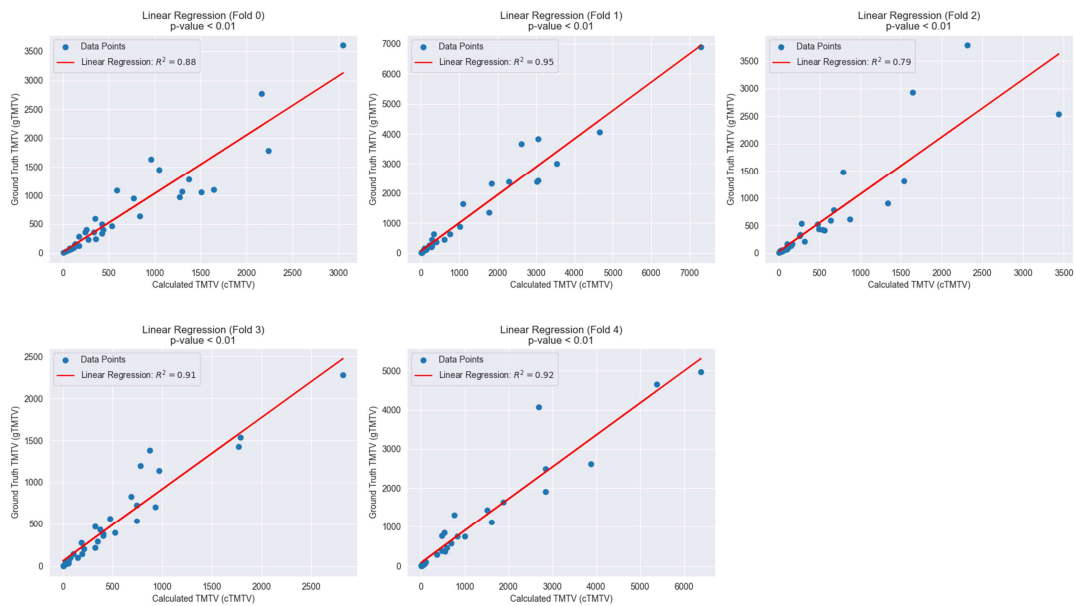

Fig. 6. Linear Regression of cTMTV vs. gTMTV. The red line represents the linear regression fit, with the coefficient of determination $R^2$ indicating the goodness of fit.

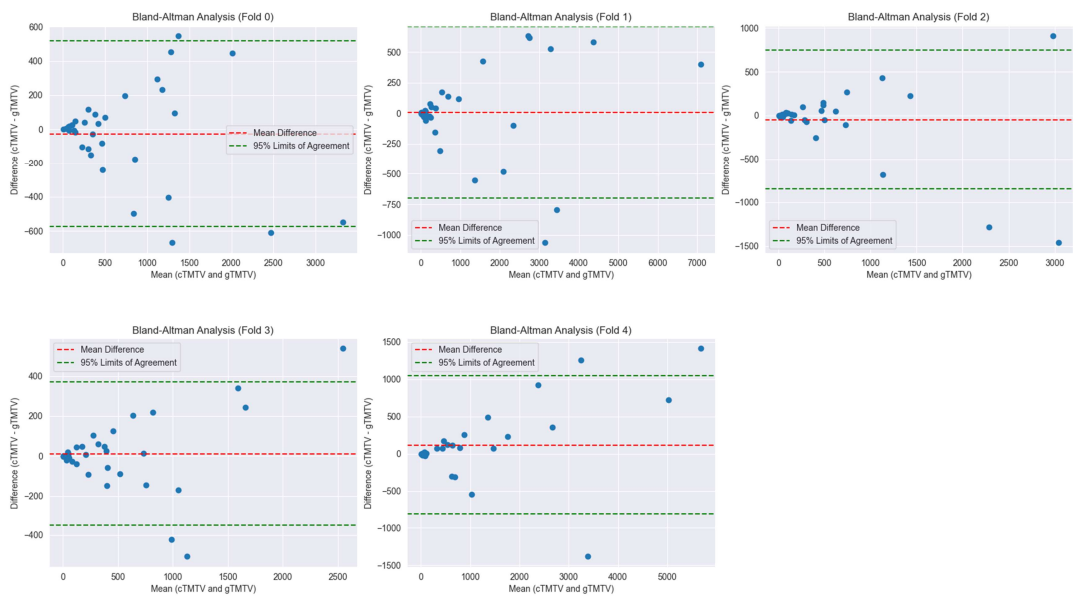

Fig. 7. Bland-Altman Analysis for cTMTV vs. gTMTV. The horizontal axis represents the mean of cTMTV and gTMTV, while the vertical axis represents their difference. The red dashed line shows the mean difference, and the green dashed lines represent the 95% limits of agreement, calculated as the mean difference ± 1.96 standard deviations of the differences.

we performed a statistical analysis of the results from five-fold cross-validation test sets. The mean cTMTV was 802.69 ± 1192.95 mL, and the mean gTMTV was 792.36 ± 1133.78 mL, resulting in a mean difference of 10.33 ± 360.03 mL. Figure 6 displays the regression analysis of cTMTV and gTMTV for each fold. The third fold had the lowest coefficient of determination $R^2$ at 0.79, which may be due to specific characteristics or anomalies in that fold's data. In contrast, the $R^2$ values for the other four folds were higher, with the second fold achieving 0.95, indicating a strong linear relationship between cTMTV and gTMTV in that fold.

Figure 7 shows the Bland-Altman analysis for each fold. Most differences fall within the acceptable range, suggesting that our calculation method generally aligns with true TMTV values. However, a few extreme points were noted, which may require further analysis and refinement to enhance consistency and accuracy.

## 4 Discussion
### 4.1 Impact of model scale

In all ablation experiments, all network parameters except the target parameters were kept constant. The default settings were as follows: each stage of the Swin Transformer had 2 layers, the patch embedding dimension was set to 24, and the number of attention heads was set to 4. To investigate the impact of different network scale parameters on segmentation performance, we adjusted the number of attention heads, Swin Transformer layers, and patch embedding dimension, while keeping other parameters fixed. The effects of these adjustments on model performance (DSC, sensitivity, precision) were evaluated.

**Table 2:** Impact of the number of attention heads.

| Number of Heads | DSC | Sensitivity | Precision |
|---|---|---|---|
| 2 | 0.7291±0.0212 | 0.7405±0.0223 | 0.7152±0.0284 |
| 4 | **0.7468±0.0171** | **0.7438±0.0099** | 0.7524±0.0112 |
| 8 | 0.7432±0.0124 | 0.7417±0.0105 | **0.7563±0.0143** |
| 16 | 0.7445±0.0131 | 0.7409±0.0111 | 0.7548±0.0127 |

We adjusted the number of attention heads to 2, 4, 8, and 16 to analyze its effect on segmentation performance (Table 2). The best DSC, sensitivity, and precision were achieved with 4 attention heads. Fewer heads (2) led to a significant performance drop, while increasing the number of heads beyond 4 slightly decreased performance. This suggests that 4 heads strike the optimal balance between feature extraction and computational complexity, effectively capturing key features of lymphoma tissue.

**Table 3:** Impact of the number of Swin Transformer layers.

| Swin Transformer layers. | DSC | Sensitivity | Precision |
|---|---|---|---|
| 2, 2, 2, 2 | 0.7455±0.0101 | **0.7538±0.0099** | **0.7584±0.0152** |
| 2, 4, 6, 8 | **0.7458±0.0098** | 0.7512±0.0086 | 0.7571±0.0135 |
| 3, 6, 9, 18 | 0.7447±0.0103 | 0.7520±0.0092 | 0.7554±0.0117 |

Next, we examined the effect of the number of Swin Transformer layers, configuring the layers to (2, 2, 2, 2), (2, 4, 6, 8), and (3, 6, 9, 18) (Table 3). Both shallow and deeper configurations performed similarly, with the shallow configuration (2, 2, 2,

2) yielding nearly identical results to deeper ones. Further increasing the number of layers resulted in a slight performance decrease, suggesting that deeper networks add complexity without significantly improving performance, potentially increasing the risk of overfitting.

**Table 4:** Impact of the Patch Embedding Dimension.

| Number of patch dimension | DSC | Sensitivity | Precision |
|---|---|---|---|
| 12 | 0.7291±0.0112 | 0.7405±0.0123 | 0.7452±0.0184 |
| 24 | **0.7353±0.0121** | **0.7479±0.0115** | 0.7458±0.0173 |
| 48 | 0.7345±0.0169 | 0.7438±0.0099 | **0.7484±0.0112** |
| 96 | 0.7289±0.0132 | 0.7443±0.0109 | 0.7415±0.0118 |

Finally, the patch embedding dimension was adjusted to 12, 24, 48, and 96 (Table 4). The best performance was observed with a dimension of 24. Smaller dimensions (12) limited the model's feature representation capacity, while larger dimensions (96) increased computational complexity and the risk of overfitting. A patch embedding dimension of 24 thus offers the optimal balance between feature representation and computational efficiency.

Based on these ablation experiments, the optimal network configuration consists of 4 attention heads, a Swin Transformer layer configuration of (2, 2, 2, 2), and a patch embedding dimension of 24. This configuration demonstrated stability across multiple experiments, significantly improving segmentation performance while maintaining low computational complexity.

### 4.2 Ablation experiment

We conducted an ablation study to evaluate the contributions of multi-scale fusion, cross-modal attention mechanisms, and inter-scale gated neural networks. The experimental setups are as follows:

- **Baseline**: The base model with single-scale fusion and dual encoders, without attention mechanisms or gated networks, serving as the benchmark.
- **MSF (Multi-Scale Fusion)**: Multi-scale fusion module is added to the baseline to assess how multi-scale features capture tumor details.
- **CMA (Cross-Modal Attention)**: A cross-modal attention mechanism is incorporated into the baseline to evaluate its role in integrating information from different modalities.
- **GFM (Gated Fusion Module)**: An inter-scale gated neural network is added to the MSF model to explore the impact of gating mechanisms on multi-scale feature fusion.
- **MSF+CMA**: Combines multi-scale fusion and cross-modal attention to evaluate their synergistic effect on segmentation performance.
- **Full Model**: The complete model integrates multi-scale fusion, cross-modal attention mechanisms, and inter-scale gated neural networks, assessing the overall impact of all combined features.

**Table 5**: Impact of each module on overall model performance.

| Model | MSF | CMA | GFM | DSC | Sensitivity | Precision |
|---|---|---|---|---|---|---|
| Baseline | ☐ | ☐ | ☐ | 0.7291±0.0112 | 0.7405±0.0123 | 0.7152±0.0184 |
| MSF | ☑ | ☐ | ☐ | 0.7386±0.0121 | 0.7549±0.0115 | 0.7228±0.0173 |

| | | | | | |
|---|---|---|---|---|---|
| CMA | ☐ | ☑ | ☐ | 0.7405±0.0101 | 0.7538±0.0099 | 0.7284±0.0112 |
| GFM | ☑ | ☐ | ☑ | 0.7369±0.0132 | 0.7443±0.0109 | 0.7415±0.0118 |
| MSF+CMA | ☑ | ☑ | ☐ | 0.7458±0.0137 | 0.7544±0.0172 | 0.7329±0.0201 |
| Full Model | ☑ | ☑ | ☑ | 0.7466±0.0118 | 0.7648±0.0145 | 0.7591±0.0178 |

Table 5 shows the impact of each module on overall model performance. The results indicate that each module contributes differently:

- **MSF**: MSF significantly improved DSC and sensitivity, confirming the importance of multi-scale information in capturing details. However, its effect on precision was more limited, suggesting the need for further refinement to reduce false positives.
- **CMA**: CMA enhanced overall performance, particularly in integrating multi-modal information. While sensitivity improved, precision saw less gain, likely due to the complexity of information fusion.
- **GFM**: GFM effectively reduced false positives and improved precision, demonstrating that gating mechanisms optimize feature selection in multi-scale fusion.
- **MSF+CMA**: Combining MSF and CMA further boosted DSC and sensitivity, indicating a synergistic effect. However, the improvement in precision remained modest, possibly due to the increased model complexity.
- **Full Model**: The complete model, integrating all key modules, achieved the best overall segmentation performance, showing that the combination of multi-scale fusion, cross-modal attention, and gated neural networks provides comprehensive optimization.

### 4.3 Limitations

While our model has demonstrated strong performance in segmenting DLBCL lesions in PET/CT images, its effectiveness is closely tied to the quality and consistency of the training data. Variations in image acquisition protocols or differences in equipment across institutions may affect the model's generalizability. Additionally, the interpretability of the model's decision-making process remains limited. Understanding how the model integrates multi-modal information for segmentation decisions is essential for clinical adoption. Future research should focus on improving model transparency, potentially through explainable AI techniques, to better support clinical decision-making.

### 5 Conclusion

In this study, we proposed a network for the automatic segmentation of DLBCL lesions in PET/CT images, which outperformed existing methods in lesion segmentation and showed robust performance in TMTV calculation. The model achieved significant improvements in both DSC and precision for segmentation, and its TMTV results confirmed its accuracy in quantifying tumor volumes. This network offers a valuable tool for lymphoma diagnosis and treatment by enhancing segmentation accuracy and TMTV estimation. Future work will focus on expanding the dataset to improve generalizability, incorporating explainable AI for better interpretability, and advancing clinical applicability to support decision-making and personalized treatment.

**Funding:** This study received support from National Natural Science Foundation of China (Grant Numbers: 62106233, 62303427, and 82370513), Young Teacher Foundation of Henan Province (Grant No.2021GGJS093), Henan Science and Technology Development Plan (Grant Number: 232102210010, 232102210062, and 222102210219) and Peking University Baidu Fund [2020BD038].